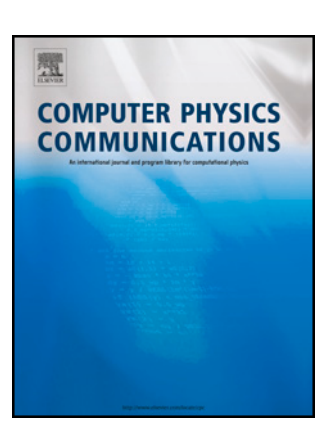

Computer Programs in Physics

# Packing3D.jl: An open-source analytical framework for computing packing density and mixing indices using partial spherical volumes

Freddie J. Barter [a,*], Christopher R.K. Windows-Yule [a,b]

[a] School of Chemical Engineering, University of Birmingham, Birmingham, B15 2TT, West Midlands, United Kingdom
[b] Positron Imaging Centre, School of Physics and Astronomy, University of Birmingham, Birmingham, B15 2TT, West Midlands, United Kingdom

ABSTRACT

Accurate quantification of local packing density and mixing in simulations of particulate systems is essential for many industrial applications. Traditional methods which simply count the number of particle centres within a given volume of space (cell) introduce discontinuities at cell boundaries, leading to unreliable measurements of packing density. We introduce Packing3D.jl, an open-source Julia package providing analytic partial-volume calculations for spheres intersecting Cartesian and cylindrical meshes. We derive closed-form solutions for single, double and triple spherical-cap intersections, plus sphere-cylinder overlaps. We implement efficient mesh-generation routines, principal-cell indexing, and data-splitting functions for time-series analyses. Performance and accuracy were validated against simple cubic and face-centred cubic lattices and via boundary-shift continuity tests. Packing3D.jl converges exactly to theoretical lattice densities, eliminates discontinuities at sub-particle resolution, and scales linearly with particle count. Memory usage remains modest (40 B per particle, 48 B per cell). Packing3D.jl provides researchers with continuous, reproducible volume-fraction fields and robust mixing indices at high performance, facilitating sensitivity analyses and optimisation in granular process engineering.

**Program summary**
*Program Title:* Packing3D.jl
*CPC Library link to program files:* https://doi.org/10.17632/srdxk6f77w.1
*Developer's repository link:* https://github.com/fjbarter/Packing3D.jl
*Licensing provisions:* MIT
*Programming language:* `Julia`
*Nature of problem:* Inaccuracy and discontinuity of packing density calculation by counting centres
*Solution method:* Derive and implement a fast algorithm for analytically calculating particle volumes intersected by planes, providing a continuous and much more accurate measurement of packing density

## 1. Introduction

Granular assemblies underpin processes such as tablet coating, vibro-packing, powder-bed fusion and silo discharge [1–3]. Powder characterisation required for optimisation of these processes frequently involves calculation of the Hausner ratio $H$ [4], defined as $H = \phi_T/\phi_B$, where $\phi_T$ and $\phi_B$ are the 'tapped' and freely settled bulk packing densities of the powder, respectively. Calculation of the Hausner ratio therefore requires measurement of packing density twice, and is dependent on this measurement being accurate and sensitive. Typically, one estimates packing density by first subdividing the domain of interest into a mesh of small volumetric elements (commonly called 'cells'). In a mesh, each cell is defined by its boundary surfaces (in Cartesian grids, these are rectangular hexahedra; in cylindrical grids, wedge-shaped cylindrical sectors). Once the mesh is in place, the total particle volume within each cell is computed and divided by the cell's volume to yield a local packing density.

In most discrete-element-method (DEM) post-processing routines, the simplest approach is applied: a particle is assigned to the single cell that contains its geometric centre, and its entire volume is counted toward that cell's total. While straightforward to implement, this 'centre-counting' rule produces step changes in packing density whenever a sphere crosses a boundary of the region, leading to inconsistent pack-ing density measurement and distorting mixing indices. Monte Carlo sampling refines this by randomly sampling points inside cell boundaries and counting those that fall within the particle volume, thereby converging in probability to the exact value, but at the cost of many samples and consequently long runtimes [5] (see Section 5.3, demonstrating a $>1000\times$ speed-up). Deterministic numerical quadrature can remove statistical noise and converge with enough samples, but still requires excessive computation for sufficient accuracy [6]. Nonetheless, numerical quadrature presents a convenient method for good estimation of volumes with complex definitions, as successfully demonstrated by Bnà et al. [7] and Chierci et al. [8] in their work on the VOFI library [9], designed for generating volume-fraction fields in fluid simulations. Analytic solutions to similar problems have also been presented, such as that by Dodd and Theodorou [10], who provided the first closed-form expressions for sphere-plane overlaps via cone-pyramid decomposition for estimation of void fraction in crystals, or the KonigCell library developed by Nicuşan and Windows-Yule [11]. However, to the best of the authors' knowledge, no existing open-source library offers a unified, closed-form solver specialised for single, double and triple spherical-cap intersections (with optional sphere-cylinder overlaps) on Cartesian or cylindrical meshes with machine-precision accuracy and microsecond-scale performance. To fill this gap, we present Packing3D.jl, a Julia package that delivers exact, continuous partial-volume calculations and standard mixing metrics (e.g. Lacey index, segregation intensity) via a simple, high-level API.

* Corresponding author.
*E-mail address:* fjb172@student.bham.ac.uk (F.J. Barter).

Building on their foundation, Packing3D.jl offers artefact-free, reproducible volume-fraction fields while remaining fast enough for large-scale DEM datasets. In the following sections, we derive the underlying geometry and benchmark our implementation against similar methods, providing researchers with a reliable, open-source tool for accurate packing and mixing analysis.

## 2. Mathematical foundations

### 2.1. Geometric primitives

We first present a method to calculate the partial volume of a sphere of radius $R$ inside an arbitrary convex cell of orthogonal planar boundaries. A sphere in a given position may intersect one, two, or three such boundaries, provided those of a given direction are separated by a distance greater than the diameter of the sphere. We denote by $a$, $b$, and $c$ the signed distances from the sphere's centre to the three orthogonal planar boundaries. For a given signed distance $d$ to a boundary, where $d \in \{a, b, c\}$, a sphere overlaps that boundary exactly when $-R < d < R$. Depending on the number of overlaps, we recognise three base geometric primitives that comprise the partial volume in question:

- **Single spherical cap.** A sphere intersecting a single plane at signed distance $a$ from its centre produces a cap of height $h = R - a$. Its volume is

$$V_{\text{cap}}(a) = \frac{\pi}{3}h^2(3R - h) = \frac{\pi}{3}(R-a)^2(2R+a) \tag{1}$$

- **Double-cap intersection.** For two orthogonal planes at signed distances $a$ and $b$ from the sphere centre, the partial volume takes the form of an intersection of two spherical caps, denoted a 'double-cap intersection.' Consider a horizontal cross-section of the sphere at height $z$. This section is a circle of radius $r = \sqrt{R^2 - z^2}$, and the area of intersection between this circle and the two overlapped boundaries (which become lines when projected onto the section) is a circular sector of angle

$$\theta_{\text{sector}} = \frac{\pi}{2} - \sin^{-1}\left(\frac{a}{r}\right) - \sin^{-1}\left(\frac{b}{r}\right) \tag{2}$$

minus two triangular areas. The triangle bases are $a$ and $b$, and the cross-sectional area therefore becomes

$$A(z) = \frac{1}{2}r^2\theta_{\text{sector}} - \frac{1}{2}a\left(\sqrt{r^2 - a^2} - b\right) - \frac{1}{2}b\left(\sqrt{r^2 - b^2} - a\right) \tag{3}$$

Integrating $A(z)$ then yields an exact closed-form solution for the double-cap intersection volume.

- **Triple-cap intersection.** When three orthogonal planes at signed distances $a, b, c$ all intersect the sphere, one again integrates the same cross-sectional formula of Equation (3), limiting the integration in $z$ so that $|z| \le R - c$. Besides the minor triple-cap intersection $(a, b, c > 0)$, each of the remaining seven sign-combinations of $(\pm a, \pm b, \pm c)$ reduces to combinations of simpler volumes, and all admit analytic antiderivatives in elementary functions.

An illustration of these three different types of analytically solved intersection volumes is given in Fig. 1. By default, sphere-cylinder intersections are simplified by assuming the cylinder's curvature is negligible compared to that of the sphere, so overlaps are computed as if the sphere intersects a plane tangent to the cylinder. Packing3D.jl does offer a numerical solution for cylindrical boundaries, but it offers minimal benefit for significant computational expense. If the user desires, it can be manually enabled with the `accurate_cylindrical` boolean keyword argument.

#### 2.1.1. Cross-section antiderivative

Upon substitution for $r$ and $\theta_{sector}$ into Equation (3), $A(z)$ becomes:

$$\begin{aligned} A(z) = &\frac{1}{4}\pi(R^2 - z^2) \\ &- \frac{1}{2}R^2\sin^{-1}\left(\frac{a}{\sqrt{R^2 - z^2}}\right) + \frac{1}{2}z^2\sin^{-1}\left(\frac{a}{\sqrt{R^2 - z^2}}\right) \\ &- \frac{1}{2}R^2\sin^{-1}\left(\frac{b}{\sqrt{R^2 - z^2}}\right) + \frac{1}{2}z^2\sin^{-1}\left(\frac{b}{\sqrt{R^2 - z^2}}\right) \\ &- \frac{1}{2}a\sqrt{R^2 - z^2 - a^2} - \frac{1}{2}b\sqrt{R^2 - z^2 - b^2} \\ &+ ab \end{aligned} \tag{4}$$

Each term can be integrated individually, which (after some rearrangement) gives:

$$\begin{aligned} \int A(z)dz = &\frac{1}{4}\pi\left(R^2 z - \frac{1}{3}z^3\right) \\ &+ \left(\frac{1}{6}z^3 - \frac{1}{2}R^2 z\right)\sin^{-1}\left(\frac{a}{\sqrt{R^2 - z^2}}\right) \\ &+ \frac{1}{3}R^3\tan^{-1}\left(\frac{az}{R\sqrt{R^2 - a^2 - z^2}}\right) \\ &+ \frac{1}{6}a(a^2 - 3R^2)\sin^{-1}\left(\frac{z}{\sqrt{R^2 - a^2}}\right) \\ &- \frac{1}{3}az\sqrt{R^2 - a^2 - z^2} \\ &+ \left(\frac{1}{6}z^3 - \frac{1}{2}R^2 z\right)\sin^{-1}\left(\frac{b}{\sqrt{R^2 - z^2}}\right) \\ &+ \frac{1}{3}R^3\tan^{-1}\left(\frac{bz}{R\sqrt{R^2 - b^2 - z^2}}\right) \\ &+ \frac{1}{6}b(b^2 - 3R^2)\sin^{-1}\left(\frac{z}{\sqrt{R^2 - b^2}}\right) \\ &- \frac{1}{3}bz\sqrt{R^2 - b^2 - z^2} \\ &+ abz + C \end{aligned} \tag{5}$$

**Fig. 1.** Schematic illustration of three cases of partial boundary overlap for a spherical particle: (**a**) single boundary intersection, (**b**) double boundary intersection, and (**c**) triple boundary intersection. The red shading indicates a partial volume of interest in each case. Intersecting planes (boundary edges) are indicated by fixed constant values for a given axis ($x = a, y = b, z = c$).

The described calculation procedure is then generalised for an arbitrary particle and an arbitrary set of boundaries; this generalisation is then used as the kernel of the package, for both the calculation of volume occupancy and concentration.

### 2.2. Continuity

Let $\phi(\mathbf{b})$ be the computed packing density when the mesh boundary is at position $\mathbf{b}$. Because each partial-volume term $V_{\text{partial}}(\delta)$ depends on the signed distance $\delta$ of the boundary from the sphere centre via smooth functions ($\sqrt{\,}$, $\sin^{-1}$, $\tan^{-1}$), it follows that

$$\lim_{\Delta \to 0} \left[ V_{\text{partial}}(\delta + \Delta) - V_{\text{partial}}(\delta) \right] = 0 \tag{6}$$

Hence, $\phi(\mathbf{b})$ is continuous in $\mathbf{b}$, in contrast to centre-counting, which jumps by $V_{\text{sphere}}$ as soon as $\mathbf{b}$ passes the centre of any sphere.

### 2.3. Complexity and performance

Each sphere-boundary overlap test and its associated analytic volume evaluation are carried out in constant time, ensuring that individual overlap calculations incur an $\mathcal{O}(1)$ cost. We first perform a series of inexpensive bounding-sphere checks: if a sphere's centre lies at least one radius inside every face of a given cell, we immediately add the sphere's full volume; if it lies more than one radius outside any face, we discard it outright; only when the centre falls within one radius of a boundary do we invoke the full analytic kernel, whether a single-cap, double-cap, or triple-cap intersection (or the sphere-cylinder formula in cylindrical coordinates). Results of practical benchmarks for the calculation in this way are provided in Section 5.

## 3. Software design and implementation

### 3.1. Package architecture

Packing3D.jl is structured around a single parent module that pulls together six submodules, each handling one core concern: meshing, geometry, I/O, utilities, and coordinate-specific workflows. The overall layout is:

- **Packing3D.jl (parent module)**. Organises all submodules, exports the public API, and defines master functions for dispatch to methods for specific coordinate systems.
- **mesh.jl**. Defines the `Mesh` type, constructs either Cartesian or cylindrical grids from user parameters or target cell counts, and provides cell-volume and cell-count metadata.
- **geometry.jl**. Implements all closed-form formulas for partial spherical volumes—single, double, and triple caps, plus sphere-cylinder overlaps—and fast inside/outside tests to short-circuit full-volume cases.
- **io.jl**. Handles data input/output and extracts particle positions and radii into Julia data structures.
- **utils.jl**. Provides shared helpers for boundary dictionary conversion, coordinate transforms, and data-splitting functions used across modules.
- **cartesian.jl**. Implements packing-density and mixing-index routines on Cartesian meshes, along with cell-indexing logic that maps $(x, y, z)$ to global cell IDs.
- **cylindrical.jl**. Mirrors the Cartesian module for $(r, \theta, z)$ meshes: packing, mixing, and cell-indexing routines tailored to cylindrical geometries.

### 3.2. Key algorithms

#### 3.2.1. Mesh generation

*Cartesian Mesh.* The Cartesian mesh is generated to produce cells of equal volume and a near-cubic aspect ratio to reduce the possibility of a particle overlapping both boundaries of a given axis. Given the boundary extents $\{x_{\min}, x_{\max}, y_{\min}, y_{\max}, z_{\min}, z_{\max}\}$ and a target total number of cells $N_{\text{target}}$, cells are created as follows:

1. Compute the ideal cubic cell side length

$$s = \left( \frac{(x_{\max} - x_{\min})(y_{\max} - y_{\min})(z_{\max} - z_{\min})}{N_{\text{target}}} \right)^{1/3} \tag{7}$$

2. Determine the number of divisions along each axis $a \in \{x, y, z\}$

$$n_a = \max\left(1, \text{round}\left(\frac{a_{\max} - a_{\min}}{s}\right)\right) \tag{8}$$

The resulting total cell count $N_{cells}$ is $n_x n_y n_z$.

3. Compute individual cell lengths per axis

$$\Delta a = \frac{(a_{\max} - a_{\min})}{n_a}, \quad a \in \{x, y, z\} \tag{9}$$

4. Generate the grid boundaries by uniform spacing

$$a_k = a_{\min} + k \Delta a, \quad k = 0, \ldots, n_a \tag{10}$$

This simple procedure ensures constant cell volume, a near-cubic shape (minimising aspect ratio), and straightforward computation of partial volumes.

*Cylindrical Mesh.* A robust method was devised for generating the cylindrical mesh, which fulfils three main criteria: Firstly, cell volume is constant to avoid bias. Secondly, it is desirable to (within reason) preserve cell shape. Thirdly, the calculation of partial particle volumes is greatly simplified if there is no point-wise convergence of cell boundaries in the centre of the cylinder. The two available approaches for the cylindrical cell generation, constant radial divisions or constant angu-lar divisions, are schematically represented in Fig. 2. Both methods are available in the package, but method (b) - constant radial divisions - is the default, as its cells have the more similar shape and radial boundary convergence at the centre is avoided. Cells were generated as follows:

1. The number of required axial divisions is calculated by rounding the cube root of the target number of cells: $n_{\text{divs},z} = \text{round}(\sqrt[3]{N_{\text{target}}})$
2. Ideal number of cells per axial division is calculated by dividing the target number of cells by the number of axial divisions: $n_{\text{cells,slice}} = \frac{N_{\text{target}}}{n_{\text{divs},z}}$
3. Number of radial divisions per axial division calculated by rounding the square root of the ideal number of cells per axial division: $n_{\text{divs},r} = \text{round}(\sqrt{n_{\text{cells,slice}}})$
4. Number of angular divisions in each radial division varies to maintain a constant cell volume

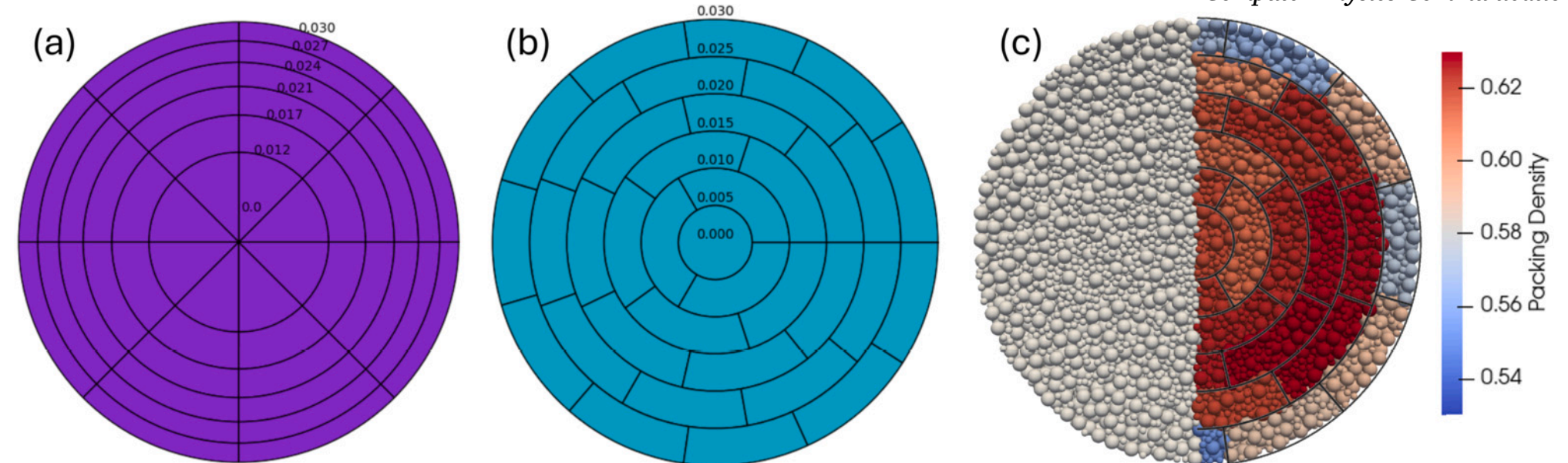

**Fig. 2.** Top-down comparative diagram of an axial division of a cylindrical mesh with radius 0.03 m, using (**a**) a constant-volume mesh preserving angular divisions but varying radial divisions, and (**b**) a constant-volume mesh preserving radial divisions but varying angular divisions, with a full cylindrical inner cell. Panel (**c**) is a half-and-half illustration of a mesh with conserved radial divisions implemented on a simulated bidisperse particle system. The right-hand half has a mesh boundary grid overlaid and its cells are coloured by local packing density.

The method for determining how the number of angular divisions should vary in each layer arises from the simple observation that the inner cell volume

$$V_{\text{inner}} = \pi(\Delta r)^2 h_{\text{slice}} \tag{11}$$

(where $h_{\text{slice}}$ is the height of a given axial division and $\Delta r$ is the radial division thickness) is exactly equal to one quarter of the total volume of the inner cell plus the second radial layer:

$$\begin{aligned} V_{\text{inner}} + V_{\text{layer 2}} &= \pi(2\Delta r)^2 h_{\text{slice}} \\ &= 4\pi(\Delta r)^2 h_{\text{slice}} \\ &= 4V_{\text{inner}} \end{aligned} \tag{12}$$

Thus, the total volume of the first radial layer must be equal to this volume minus the volume of the inner cell:

$$V_{\text{layer 2}} = 4V_{\text{inner}} - V_{\text{inner}} = 3V_{\text{inner}} \tag{13}$$

This pattern repeats, with the third layer containing $5V_{inner}$, the fourth containing $7V_{inner}$, and so on until $(2n_{divs,r} - 1)V_{inner}$ for the outer layer. This is used directly to divide each radial layer. The total number of cells in a given slice is therefore equal to:

$$N_{\text{slice}} = \sum_{n=1}^{n_{\text{divs},r}} (2n-1) = (n_{\text{divs},r})^2 \tag{14}$$

This result is the reason for the square root in step 3, with the rounded cube root in step 1 designed to mirror the division process for that of a cubic region.

### 3.2.2. Packing density calculation

Packing density in a user-defined region is computed as the ratio of total particle volume to region volume. This proceeds in three steps:

1. **Define the region**. The user supplies explicit boundary limits for either a Cartesian region $\{x_{\min}, x_{\max}, y_{\min}, y_{\max}, z_{\min}, z_{\max}\}$ or a cylindrical region $\{r_{\min}, r_{\max}, \theta_{\min}, \theta_{\max}, z_{\min}, z_{\max}\}$.
2. **Accumulate particle volumes**. For each particle, we determine which cell(s) it overlaps using the cell-indexing routines. Before any detailed calculation, a quick "all-inside" or "all-outside" test either adds the full sphere volume or skips the particle. Only if the centre lies within one radius of a face do we call the analytic partial-volume kernels (single, double, triple cap or sphere-cylinder). These kernels return the exact volume of the portion of the sphere inside the cell.
3. **Compute packing density**. Summing all contributions yields the total particle volume in the region. The packing density is then simply

$$\phi = \frac{(\text{total particle volume})}{(\text{region volume})}, \tag{15}$$

a single Float64 value that faithfully reflects partial overlaps at every boundary.

### 3.2.3. Dataset partitioning

The mixing indices available in Packing3D.jl require the partitioning of a single particle dataset into two mutually exclusive populations, which must be consistently tracked across multiple simulation frames. To facilitate this, Packing3D.jl provides two complementary public functions: `split_data` and `match_split_data`. The workflow proceeds in two stages:

*Initial Split.* The function `split_data` performs a one-off partition of a data dictionary into two subsets, based on either spatial coordinates or intrinsic particle properties such as radius, density, or type. Depending on the criteria, it uses a median or explicit threshold, or tolerance-based value matching, to assign each point ID to one of two groups. Crucially, it returns two `Set{Int}` objects containing unique particle IDs. Julia's `Set` type is unordered, prohibits duplicates, and provides hashed $\mathcal{O}(1)$ average-time membership tests, making it an ideal candidate to store the IDs of fixed particle populations.

*Frame-by-Frame Matching.* Once the initial partition is established, `match_split_data` applies the same classification to subsequent data frames. It accepts a new data dictionary and the two precomputed ID sets, constructs Boolean masks by testing each point ID for membership in the sets, and then invokes the shared helper `extract_points` to materialise the two subsets. Because particle IDs remain invariant across frames, no re-evaluation of splitting criteria is necessary (only fast hash-based lookups) making this approach ideal for time-series analyses. Separating the expensive, metadata-rich splitting operation from the lightweight, high-throughput matching step minimises redundant computation and preserving the overall linear scaling of the packing and mixing kernels.

### 3.2.4. Lacey mixing index

The Lacey mixing index $M$ provides a normalised measure of how well two species are intermixed, defined by

$$M = \frac{\sigma^2_{\text{segregated}} - \sigma^2_{\text{actual}}}{\sigma^2_{\text{segregated}} - \sigma^2_{\text{random}}} \tag{16}$$

where $\sigma^2_{\text{actual}}$ is the measured variance in species concentration $c_i$ across cells, $\sigma^2_{\text{random}} = c^*(1-c^*)/N$ is the variance of a perfectly random mixture, and $\sigma^2_{\text{segregated}} = c^*(1-c^*)$ is the variance of a perfectly segregated state [12]. $N$ is the average number of particles per cell, $c^*$ is the bulk concentration, and $V_{1,i}$ and $V_{2,i}$ are the volumes occupied by species 1 and 2 in cell $i$, respectively. $c_i$ and $c^*$ are defined as:

$$c_i = \frac{V_{1,i}}{V_{1,i} + V_{2,i}}, \quad c^* = \frac{\sum_i V_{1,i}}{\sum_i (V_{1,i} + V_{2,i})} \tag{17}$$

The implemented procedure for calculating the Lacey mixing index is as follows:

1. **Mesh generation**. From user-provided region limits and a target cell count (or explicit divisions), a `Mesh` object is constructed in either Cartesian or cylindrical coordinates, as described in Section 3.2.1. This mesh defines the constant-volume cells over which both species' volumes will be evaluated.
2. **Per-cell volume kernel**. We then call `compute_volumes_per_cell()`, which loops over all particles in each dataset and implements:
   - **Principal-cell assignment**. Each particle's centre is mapped to a 'principal' cell via fast cell-indexing logic where possible. Natively generated `Mesh` structures are geometrically defined in such a way that the particle's principal cell index can be evaluated purely from its position, avoiding $\mathcal{O}(N_{\text{cells}})$ computation and offering excellent efficiency. If the particle centre lies at least one radius inside all faces of its principal cell, its full volume is assigned there.
   - **Overlap detection and partial volume distribution**. If the centre of the sphere lies within one a distance of one radius from a boundary, we inspect which faces (up to three) the sphere intersects, gathering the signed overlap distances. Then, the analytic kernel for the principal overlap case (single, double, or triple cap) yields the volume within the principal cell. Adjacent cells receive the remaining volume sequentially, by evaluating simpler intersections (e.g. single-cap on the next face) where possible, for each overlapped direction and subtracting previously assigned portions. This technique ensures that the sum of contributions exactly equals the full sphere volume, as demonstrated in Section 4.3.
   - **Population volume accumulation**. The per-cell volume of each population is then accumulated in two Float64 arrays, which is then readily used to calculate either packing density or concentration.
3. **Concentration and Variance**. For each cell $i$, the local concentration is weighted by cell occupancy to compute the measured variance

$$\sigma^2_{\text{actual}} = \frac{\sum (V_{1,i} + V_{2,i})(c_i - c^*)^2}{\sum (V_{1,i} + V_{2,i})} \tag{18}$$

   The Lacey Mixing Index is then calculated by substitution into Equation (16).

### 3.3. Usage examples

Below are three representative workflows using Packing3D.jl. In each case the package dispatches automatically to Cartesian or cylindrical routines based on the `system` keyword. Interactive examples can be found and adjusted by prospective users in the Jupyter Notebook available using the `launch binder` badge in the GitHub repository: https://github.com/fjbarter/Packing3D.jl.

#### 3.3.1. Packing density in a cylindrical region

Compute the local packing density of spherical particles in a full cylindrical region.

```julia
using Packing3D

data = read_vtk_file("particles.vtk")

packing_density = calculate_packing(
    data;
    boundaries = Dict(
      :r_min => -1, # <0, full circle
      :r_max => 0.0375,
      :theta_min => 0.0,
      :theta_max => 2pi,
      :z_min => 0.005,
      :z_max => 0.020
    ),
    system = :cylindrical
)

println("Packing density: ", packing_density)
```

#### 3.3.2. Lacey mixing index in Cartesian coordinates

Quantify the degree of mixing between two particle populations on a uniform Cartesian grid, splitting data by radius.

```julia
using Packing3D

# load two datasets from VTK
data = read_vtk_file("particles.vtk")

# generate id Sets for data partitions
data_1_ids, data_2_ids = split_data(
    data;
    split_by=:radius,
    threshold=0.0007
)

# match id Sets to the original dataset
data1, data2 = match_split_data(
    data,
    data_1_ids,
    data_2_ids
)

lacey_index = calculate_lacey(
    data1, data2;
    system = :cartesian,
    boundaries = Dict(
      :x_min => -0.05, :x_max =>  0.05,
      :y_min => -0.05, :y_max =>  0.05,
      :z_min =>  0.00, :z_max =>  0.10
    ),
    target_num_cells = 1000,
    calculate_partial_volumes = true
)

println("Lacey index: ", lacey_index)
```

#### 3.3.3. Inspecting mesh properties

Generate a mesh explicitly and query its cell count and volume.

```julia
using Packing3D

mesh = Mesh(
    :cylindrical;
    divisions = Dict(:r=>5, :theta=>3, :z=>8),
    params = Dict(
      :cylinder_radius    => 0.03,
      :cylinder_base_level => 0.0,
      :cylinder_height    => 0.08
    )
```

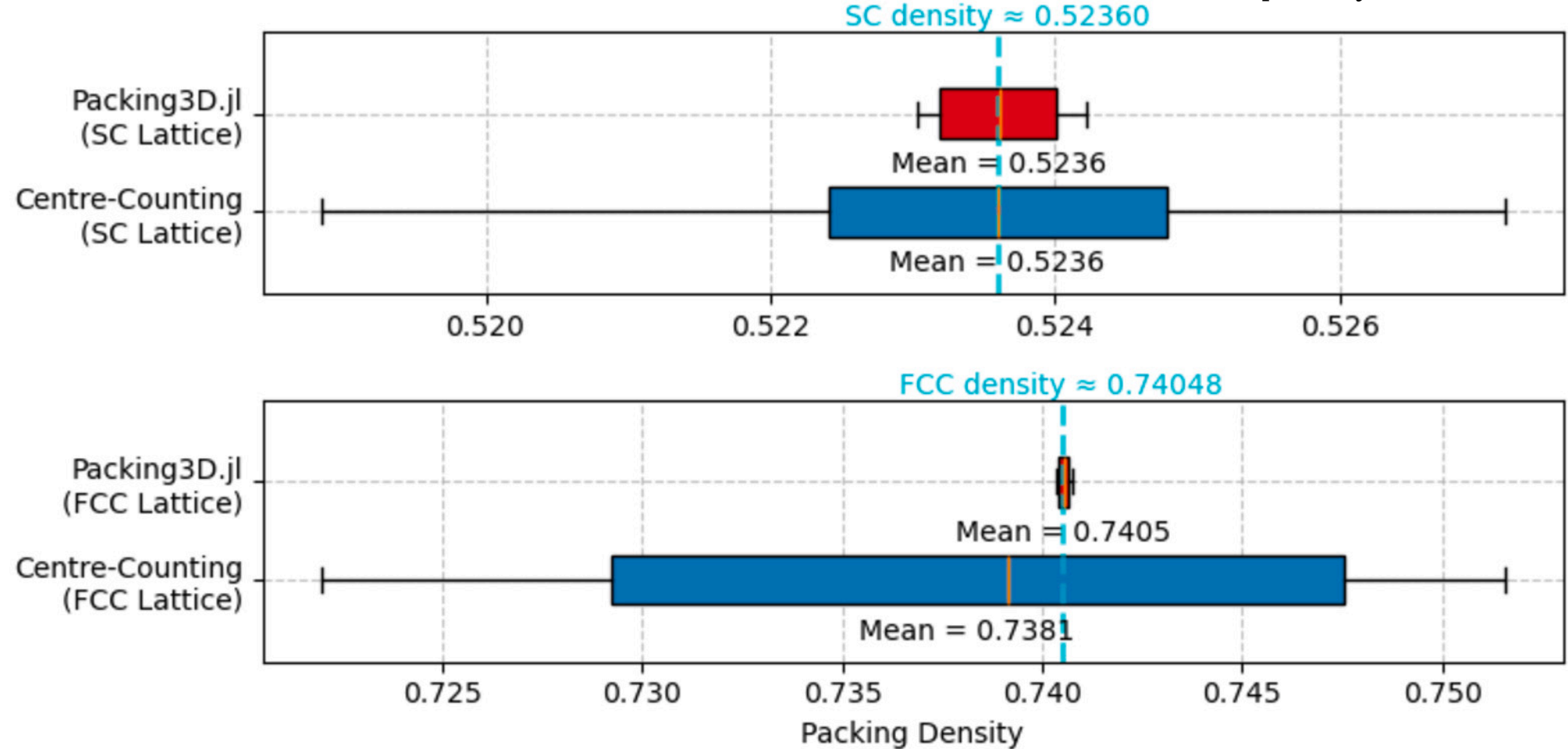

**Fig. 3.** Boxplots comparing packing-density measurements in simple cubic (SC) and face-centred cubic (FCC) lattices (containing roughly 100 000 monodisperse spheres) using centre-counting (blue) and Packing3D.jl's analytic method (red). In each plot, the central orange line is the median of 1000 trials (region translated by one small sphere radius), the box spans the 25th–75th percentiles and whiskers extend to 1.5× IQR. The dashed horizontal line marks the exact theoretical value (SC: $\pi/6 \approx 0.52360$; FCC: $\pi\sqrt{2}/6 \approx 0.74048$).

```
)

println("Total cells: ", get_total_cells(mesh))
println("Cell volume: ", get_cell_volume(mesh))
```

## 4. Validation

The accuracy, robustness and practical utility of Packing3D.jl were evaluated through two comprehensive test programmes that probe both idealised and industrially relevant conditions. Firstly, a test was conducted to assess the accuracy and consistency of Packing3D.jl in measuring the analytically known packing density of lattice structures. Secondly, the continuity of its measurement was verified by running a translational boundary test, in which a Cartesian region is moved throughout a single particle radius. In both tests, Packing3D.jl's analytical algorithm was compared to an equivalent centre-counting one.

### 4.1. Analytical lattice benchmarks

We first verified the implementation against two crystalline packings for which the exact packing density is known analytically: simple cubic (SC) with $\phi_{SC} = \pi/6 \approx 0.52360$ and face-centred cubic (FCC) with $\phi_{FCC} = \pi\sqrt{2}/6 \approx 0.74048$. An SC lattice of 125 000 spheres was generated, and Cartesian and cylindrical regions selected such that the majority of the spheres was included in each. This region was randomly translated by $\Delta \in [-r_{min}, r_{min}]$ 1000 times, with the packing density calculated each time, to generate a distribution of measured values for both centre-counting and the analytic formulation offered by Packing3D.jl. This procedure was then repeated for an FCC lattice of 108 000 spheres. The results of this investigation are shown in Fig. 3. Packing3D.jl obtained a highly accurate measurement of both lattices ($\Delta\phi < 0.001$) and its standard deviation was approximately 8 times smaller with the SC lattice and approximately 40 times smaller with the FCC lattice.

### 4.2. Continuity and smoothness

To verify that Packing3D.jl truly eliminates the discontinuities inherent in centre-counting, we performed a two-part boundary-shift experiment on a mixed bidisperse bed, comprised of small (1 mm diameter) and large (2 mm diameter) particles in a cylindrical vessel (60 mm diameter). The analysis region was translated in the $x$-direction over $N = 1000$ evenly spaced positions, and the packing density was computed both by simple centre-counting and by the analytic partial-volume method of Packing3D.jl.

In the first, large-amplitude sweep we moved the boundary over $\Delta x \in [0, 0.01]$ m, corresponding to translation through 10 small-particle diameters, designed to capture genuine changes in packing density throughout the bed. As expected, the centre-counting curve (blue line in Fig. 4a) exhibits abrupt jumps and varies significantly. By contrast, the analytic method produces a perfectly continuous response, with no visible discontinuities and a maximum local slope of $\left|\frac{d\phi}{db}\right|_{max} \approx 0.00347\ r_{small}^{-1}$.

In the second, fine-scale sweep the boundary moved over $\Delta x \in [0, 0.0005]$ m, corresponding to movement through one small-particle radius. Here the analytic packing density varied by less than 0.001 across the entire interval and exhibited a root-mean-square boundary derivative of $\left|\frac{d\phi}{db}\right|_{RMS} \approx 2.147 \times 10^{-7}\ r_{small}^{-1}$, confirming finiteness and stability at sub-particle resolution (Fig. 4b). The centre-counting result, by contrast, exhibits discrete small and large jumps of 0.000037 and 0.0003, corresponding to the inclusion or exclusion of small or large particles, respectively.

### 4.3. Volume conservation

To verify that no volume is lost or double-counted when partitioning spheres into cells, we applied a 1000-cell Cartesian mesh that fully encapsulated each of five simple-cubic lattices ranging in size from $92^3$ to $171^3$ spheres, and extracted the computed particle volumes in each cell. We then evaluated

$$\Delta V = \sum_{\text{cells}} V_{\text{cell}} - \sum_{\text{particles}} V_{\text{particle}}, \tag{19}$$

where $V_{cell}$ is the total volume occupied by particles in a given cell and $V_{particle}$ is the volume of a given particle. Shown in Table 1, the absolute error $\Delta V$ never exceeds $5 \times 10^{-11}$ m$^3$ (remaining below 1/5000th the volume of a single particle) and the relative error stays below $4 \times 10^{-11}$, confirming the presence of only accumulated machine-precision error.

These experiments demonstrate that the analytic partial-volume kernel in Packing3D.jl yields a packing-density field that is not only continuous but also differentiable, even at resolutions well below the particle size, making it ideally suited for any application requiring smooth spatial gradients or sensitivity analyses.

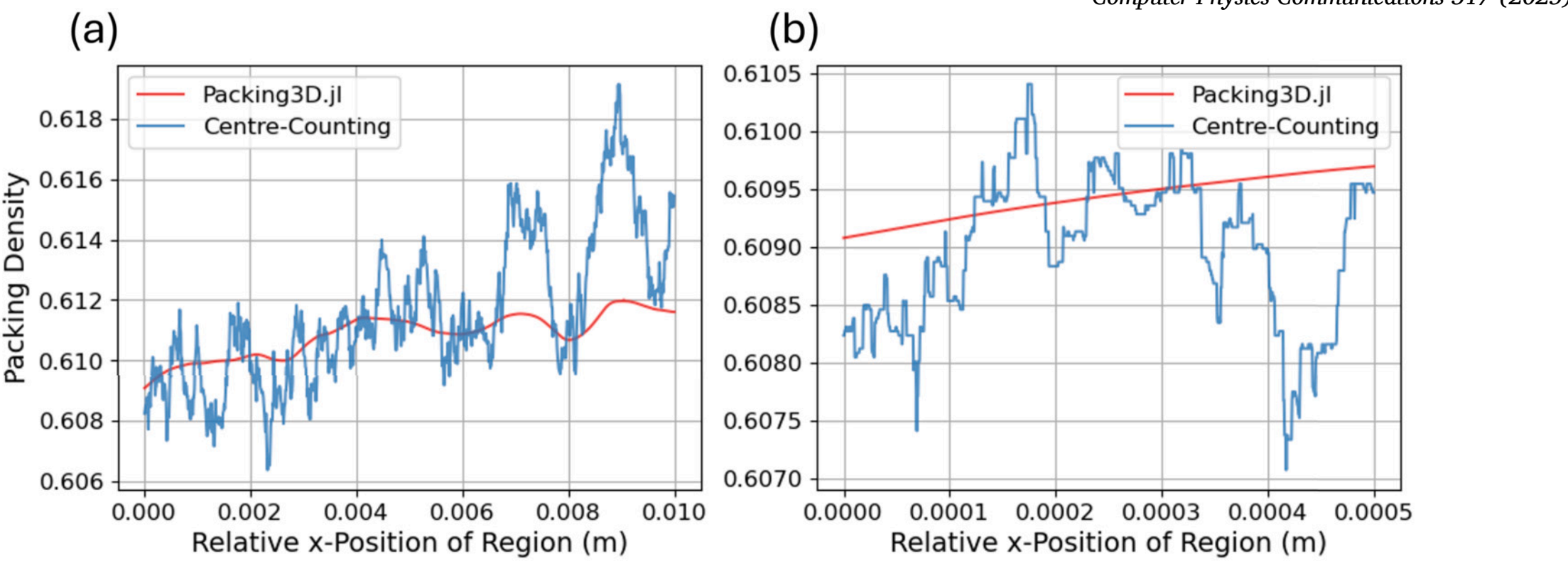

**Fig. 4.** Boundary shift test for packing density in Cartesian coordinates. (**a**) Large-amplitude sweep (0 – 0.01 m): centre-counting (blue) shows discrete jumps, while the analytic method of Packing3D.jl (red) is continuous. (**b**) Fine-scale sweep (0 – 0.0005 m): the analytic response remains smooth and differentiable.

**Table 1**
Volume conservation for various lattice sizes. $V_{\text{particle}} \approx 2.681 \times 10^{-7}\ \text{m}^3$.

| Number of spheres | $\Delta V$ ($\text{m}^3$) | Relative error |
|---|---|---|
| 778 688 ($92^3$) | $-3.40 \times 10^{-12}$ | $1.63 \times 10^{-11}$ |
| 1 860 867 ($123^3$) | $1.54 \times 10^{-11}$ | $3.08 \times 10^{-11}$ |
| 2 924 207 ($143^3$) | $-2.06 \times 10^{-11}$ | $2.63 \times 10^{-11}$ |
| 3 944 312 ($158^3$) | $-3.16 \times 10^{-11}$ | $2.99 \times 10^{-11}$ |
| 5 000 211 ($171^3$) | $4.93 \times 10^{-11}$ | $3.68 \times 10^{-11}$ |

## 5. Performance benchmarking

Practical utility requires not only numerical accuracy but also sufficient throughput capability, especially when modern DEM simulations routinely track millions of particles. Accordingly, we benchmarked the speed of Packing3D.jl as a function of particle count. All tests were executed on a Lenovo IdeaPad Flex 5 (11th Gen Intel Core i5-1135G7 @ 2.4 GHz, 8 GB RAM) running Windows 10 Home, using Julia 1.11 in single-threaded mode. Each data point is the mean of 10 repeats. We also provide a pseudocode summary for the naïve centre-counting algorithm, as well as for the two alternative convergent methods.

### 5.1. Pseudocode for alternative algorithms

Listing 1: Centre-Counting volume estimation

```
Inputs:  sphere centre (x, y, z),
         sphere radius R

function CentreCounting(R, x, y, z):
  if x_min ≤ x ≤ x_max and
     y_min ≤ y ≤ y_max and
     z_min ≤ z ≤ z_max then
    return (4/3)πR^3    # Entire sphere counted
  else
    return 0           # Volume discarded
  end
end
```

Listing 2: Monte Carlo volume estimation

```
Inputs:  sphere radius R,
         cell planes at distances a, b, c,
         samples N

function MonteCarlo(R, a, b, c, N):
  count_in = 0
  for i in 1 : N:
    # Sample uniformly in [-R,R]^3
    (x, y, z) ~ Uniform([-R, R]^3)
    # Test whether point lies in the intersection
    if x > a and y > b and z > c and x^2 + y^2 + z^2 ≤ R^2:
        count_in += 1
  frac = count_in / N
  return frac * 8 * R^3
```

Listing 3: Two-stage Simpson quadrature

```
Inputs:  sphere radius R,
         cell planes at distances a, b, c,
         number of samples N_z, N_x

Calls:   Simpson1D # simple helper

function SimpsonQuadrature(R, a, b, c, N_z, N_x):
  # Dispatch to edge case handlers if needed
  # Stage 1: compute A(z) values
  z_min = max(-R, c);    z_max = R
  Δz = (z_max - z_min)/(N_z - 1)
  for j in 0 : N_z - 1:
    z = z_min + j Δz
    r = sqrt(R^2 - z^2)
    x_min = max(-r, a);    x_max = r
    Δx = (x_max - x_min)/(N_x - 1)
    for i in 0 : N_x - 1:
      x = x_min + i Δx
      y_high = sqrt(r^2 - x^2)
      y_low = max(b, -y_high)
      f_i = max(0, y_high - y_low)
    A_j = Simpson1D(f; Δx)
  # Stage 2: integrate A(z) over z
  return Simpson1D(A; Δz)
```

### 5.2. Particle throughput benchmarking

To quantify the raw cost of our analytic partial-volume packing-density kernel, we processed a series of monodisperse simple-cubic lattice files containing $N$ particles each. In the Cartesian tests, each dataset was loaded and the region of interest defined by setting the $x$, $y$, and $z$ bounds to $\pm r\sqrt[3]{N}$. Packing density was then computed twice for each case: once with the full analytic partial-volume logic and once with centre-counting. Wall-times were recorded via BenchmarkTools' `@benchmark` macro over ten samples and averaged, showing a linear increase in compute time with $N$.

An equivalent set of benchmarks was carried out in a cylindrical region by replacing the Cartesian bounds with radial limits equal to the lattice half-width, full angular coverage, and identical axial extents. Again, packing density was timed in both analytic and centre-counting modes. The resulting performance response (Fig. 5) is linear in $N$, as expected. The analytic partial-volume algorithm will never be the rate-limiting step in a DEM workflow. Its cost grows linearly with system size and remains on the order of seconds for realistic simulation sizes: a dataset of $5 \times 10^6$ spheres has packing density calculated in a maximum of roughly 0.8 s, and Lacey index calculated in under 4 s. Regardless of system size, the post-processing calculation time will be negligible beside the many thousands of seconds consumed by the simulation itself. These benchmarks exclude file-loading overhead, whose duration depends entirely on the storage medium: when the analysis is integrated directly into the simulation no loading is needed, whereas reading legacy ASCII VTK output can take orders of magnitude longer for large files. In practice, therefore, the analytic routine adds no perceptible bottleneck while simultaneously eliminating the discontinuities inherent in centre-counting and furnishing smooth, physically meaningful packing-density fields.

**Fig. 5.** Benchmark results for (**a**, **b**) packing density and (**c**, **d**) Lacey index versus number of particles. Both Packing3D.jl's analytical algorithm (red squares) and a centre-counting method (blue triangles) were run ten times for each data point; the plotted value is the mean wall time, with error bars indicating one standard deviation across the ten runs.

### 5.3. Comparison to Monte Carlo and numerical quadrature

For a fair comparison of the convergent volume calculation methods with no bias towards a particular system, we extended our benchmarking to 100 random configurations of $(a, b, c) \in [-0.9R, 0.9R]^3$, each determining the volume of a unique triple-cap intersection ($R = 1.0$), comparing: Monte Carlo sampling, a numerical quadrature approach, and our analytic kernel. The quadrature method uses a two-stage composite Simpson's rule, selected to balance typical implementation and computational efficiency, expected to be a fair representation of other quadrature-based methods, such as that developed by Bnà et al. [7] or Lopez [13] in volume-of-fluid field estimation software. First, we sample the cross-sectional intersection area $A(z)$ at evenly spaced $z$-levels, calculated using Simpson's rule in $x$, and then we apply Simpson's rule again to integrate those slice areas over $z$ to recover the total volume. This two-stage method, therefore, is effectively numerical integration of equation (4). We imposed a 1% maximum tolerable relative error and then tuned each method to meet it. For the nested Simpson quadrature this meant using 51 samples per stage, producing an absolute error of roughly $5 \times 10^{-3}$ in our test, while Monte Carlo sampling required $10^5$ trials to achieve the same error bound. We then recorded for each approach the mean absolute error $\bar{\varepsilon} = |V_{\text{approx}} - V_{\text{exact}}|$ and the mean wall-time $\bar{t}$. All benchmarks were performed on the same test machine as in Section 5.2, and the results are summarised in Table 2.

Because our analytic solver is built on exact, closed-form antiderivatives, expressed in a finite sequence of elementary arithmetic and trigonometric operations, it suffers only standard IEEE-754 [14] floating-point round-off (on the order of $10^{-15}$ relative error). In practice this means the analytic volumes agree with the true values to within a few units of machine epsilon in every case.

Monte Carlo sampling is orders of magnitude too slow to be practical for large-scale or real-time analyses. Simpson quadrature improves performance significantly, reducing wall time by more than 20×, but the nested integration logic is substantially more complex to implement. By contrast, Packing3D.jl's analytic kernel delivers machine-precision accuracy in just 1.87 μs, outperforming quadrature by >40× and Monte Carlo by >1000×. Although deriving and coding the closed-form expressions demanded significant theoretical and development effort, the final implementation is compact, robust, and yields unparalleled speed and precision for partial-volume calculations.

### 5.4. Memory footprint

Because calculations are streaming and overlap tests are stateless, memory usage is dominated by (i) particle arrays and (ii) mesh cell information. Particles will generally have 5 associated 64-bit values necessary for processing with Packing3D.jl (3D position, radius, ID), for a total of 40 B per particle. Other data can be loaded but is unnecessary and should be ignored if dealing with a very large dataset. Cells require 6 Float64 values each (2 boundaries per dimension), totalling 48 B per cell. There is then a fixed $\mathcal{O}(1)$ memory requirement for organisation. The full analytic pipeline therefore requires only

$$\text{RAM} \approx [40 N_{\text{particles}} + 48 N_{\text{cells}} + \mathcal{O}(1)]\,\text{B} \tag{20}$$

with an asymptotic space complexity in $\mathcal{O}(N_{\text{particles}})$, assuming $N_{\text{particles}} \gg N_{\text{cells}}$. Loading and processing 100 million particles therefore requires approximately 4 GB of RAM, easily fitting on a commodity workstation.

## 6. Conclusion

We have presented Packing3D.jl, an open-source Julia library that implements analytic partial-volume calculations for spheres intersecting arbitrary Cartesian and cylindrical meshes. By extending the classic sphere-plane solution to handle single, double and triple cap intersections, as well as sphere-cylinder overlaps, Packing3D.jl delivers continuous, artefact-free packing densities and mixing indices (e.g. Lacey index) without the jumps inherent to centre-counting and at a fraction of the cost of Monte Carlo sampling. Our validation against simple cubic and face-centred cubic lattices confirmed exact convergence to theoretical values, while boundary-shift experiments demonstrated smooth, differentiable behaviour even at sub-particle resolution. Benchmarking on datasets up to millions of particles showed linear scaling and throughputs on the order of $10^8$ processed particles per second in single-threaded Julia, with modest memory requirements (40 B per particle, 48 B per cell). In addition, a comprehensive benchmark against Monte Carlo sampling and nested Simpson quadrature has shown Packing3D.jl to be over 1000× and 40× faster, respectively, while delivering machine-precision accuracy. Volume-conservation tests on simple-cubic lattices up to 5 million spheres confirmed only floating-point round-off error, demonstrating exact partitioning across cells.

**Table 2**
Mean absolute error and mean wall-time over 100 random triple-cap tests. Average analytic volume $\approx 0.5731$.

| Method | $\bar{\varepsilon}$ | $\bar{t}$ (ms) | Analytic speed-up ($\times$) |
|---|---|---|---|
| Monte Carlo ($N = 10^5$) | 0.00221 | 2.26 | 1210 |
| Simpson quadrature ($51 \times 51$ pts) | 0.00472 | 0.0849 | 45.3 |
| **Analytic (Packing3D.jl)** | $< 10^{-14}$ | 0.00187 | - |

Beyond its core packing-density and mixing-index routines, Packing3D.jl offers flexible mesh generation, efficient data-splitting across frames, and seamless support for both Cartesian and cylindrical coordinate systems via a unified API. Its high accuracy, performance, and ease of integration make it a powerful tool for DEM post-processing. Future work will focus on extension to non-spherical shapes (e.g. ellipsoids, polyhedra), automatic parallelisation, and coupling with simulations to provide real-time analysis. By providing researchers with smooth, reproducible volume-fraction fields and robust mixing metrics, Packing3D.jl will help pave the way for more reliable sensitivity analyses and optimisation in granular process engineering.

## CRediT authorship contribution statement

**Freddie J. Barter:** Writing – original draft, Visualization, Validation, Software, Methodology, Investigation, Formal analysis, Data curation, Conceptualization. **Christopher R.K. Windows-Yule:** Writing – review & editing, Supervision.

## Declaration of competing interest

The authors declare that they have no known competing financial interests or personal relationships that could have appeared to influence the work reported in this paper.

## Data availability

Data will be made available on request.